\documentclass[aps,pra,twocolumn,groupedaddress]{revtex4}

\usepackage{graphicx}
\usepackage{dcolumn}
\usepackage{color}
\usepackage{bm}
\usepackage{amsmath}
\usepackage{dsfont}

\begin{document}

\title{Spatial modes in waveguided parametric down-conversion}

\author{Andreas Christ}
\email[]{andreas.christ@mpl.mpg.de}
\homepage[]{http://mpl.mpg.de}
\author{Kaisa Laiho}
\author{Andreas Eckstein}
\author{Thomas Lauckner}
\author{Peter J. Mosley}
\author{Christine Silberhorn}
\affiliation{Max Planck Institute for the Science of Light, 
G\"unther-Scharowsky-Str. 1/Building 24, 91058 Erlangen, Germany}


\begin{abstract}
The propagation of several spatial modes has a significant impact on the
structure of the emission from parametric down-conversion in a nonlinear
waveguide. This manifests itself not only in the spatial correlations of the photon pairs 
but also, due to new phase-matching conditions, in the output spectrum, radically altering
the degree of entanglement within each pair. Here we investigate both theoretically
and experimentally the results of higher-order spatial-mode propagation in nonlinear waveguides.
We derive conditions for the creation of pairs in these modes and present
observations of higher-order mode propagation in both the spatial and spectral domains. Furthermore, we observe correlations between the different   
degrees of freedom and finally
we discuss strategies for mitigating any detrimental effects and optimizing pair production in the fundamental mode.

\end{abstract}

\maketitle


\section{Introduction}
The process of parametric down-conversion (PDC) is widely used in quantum optics and quantum information as a source of photon pairs. These photon pairs are generated inside a nonlinear crystal when a pump photon decays into signal and idler photons; typically the signal photon is used to herald the availability of an idler photon that can be used in subsequent experiments.  The distribution of the pump photon energy between the daughter photons is determined by momentum conservation or, using the language common in optics, by the phase-matching conditions.

After some initial discussion, experiments have demonstrated that for PDC in bulk crystals the orbital angular momentum (OAM) of the pump beam is conserved \cite{arlt_parametric_1999,mair_entanglement_2001}. Hence, by controlling the spatial mode of the interacting beams a new degree of freedom becomes available for quantum information processing. It can be used to generate mode-entangled \cite{arnaut_orbital_2000,mair_entanglement_2001} and even hyperentangled bi-photon states for quantum metrology \cite{walborn_hyperentanglement-assisted_2003} or quantum information-processing applications \cite{torres_preparation_2003, garcia-escartin_quantum_2008}.

In recent years attention has fallen on PDC in waveguiding structures \cite{chen_versatile_2009, anderson_quadrature_1995, uren_efficient_2004, tanzilli_highly_2001}. The resulting implicit spatial-mode control has significant benefits for photon-pair production. Due to the tight confinement of the fields inside the waveguide (WG) the effective down-conversion rate into useful spatial modes increases substantially. This, in turn, allows such sources to be pumped at greatly reduced power levels while still achieving high photon fluxes relative to bulk crystal downconverters. In the ideal case only the  fundamental spatial mode is present, and  the waveguide output is naturally suited for efficient coupling to fibers yielding a robust, high-brightness photon-pair source.

However, real-world waveguide sources usually deviate significantly from this ideal. Typical waveguides support the propagation of several spatial modes for the interacting pump, signal, and idler photons which can assume any of the spatial modes guided at their respective wavelengths \cite{banaszek_generation_2001}. In the standard case of a roughly rectangular waveguide, the radial symmetry of the system is broken and hence the OAM conservation condition is reduced to that of parity conservation between the spatial modes of the three fields. Therefore OAM is no longer a useful parameter. Instead the fields must be decomposed into the waveguide transverse field mode solutions.

In addition to the modification of the spatial characteristics introduced by the waveguide, the spectral properties of the PDC process are also affected. The energy distribution between the signal and idler photons, and hence their wavelengths, is governed by the phase-matching of the longitudinal components of the wave vector. Every spatial mode corresponds to a different transverse --- and therefore also longitudinal --- momentum, so for every set of three spatial modes of pump, signal, and idler we observe photon pairs with a specific distribution of wavelengths. The resulting output from the waveguide contains many of these distributions coherently superimposed on one another. For various applications this can be a significant problem if a heralded single photon in a single well-defined spatio-temporal mode is needed. Likewise second-harmonic generation (SHG), the reverse process of PDC, is affected.

In this paper, we present a study of the impact of discrete spatial mode structure on PDC in a nonlinear waveguide. We consider the connection between the spatial and spectral degrees of freedom  as well as both the challenges and benefits that arise from using a waveguided PDC photon-pair source in quantum information processing. In Secs. II and III we present the theoretical groundwork for dealing with discrete spatial modes in PDC. In Sec. IV we investigate the spatio-spectral mode structure of our waveguide utilizing SHG. We present experimental results on the spectral impact of spatial multi-mode PDC and theoretical predictions in Sec. V. Section VI discusses our findings with respect to spatio-spectral correlations.

\section{Theoretical background}
We analyze the process of PDC in the interaction picture. In order to derive  the two-photon state of the down-converted fields we follow an approach used in \cite{fiorentino_spontaneous_2007, grice_spectral_1997, Mikhailova2008}.
\begin{figure}
\includegraphics[width=0.6\linewidth]{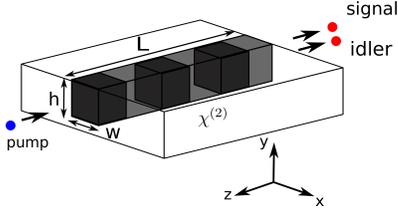}
\caption{\label{fig:waveguideSchematic} (Color online) Schematic picture of waveguided PDC in a periodically poled rectangular
waveguide.}
\end{figure}
Assuming a nonlinear dielectric crystal with length $L$, a constant waveguide cross section $A$, and propagation along the $z$ direction (Fig. \ref{fig:waveguideSchematic}), the down-converted bi-photon quantum state \(\left|\psi\right>\) (within first-order perturbation theory) is then given by:
\begin{equation}
     \left|\psi\right> \sim  \int_0^{t'}\mathrm
    dt\,\int_V \mathrm d V \, \chi^{\left(2\right)}\left(z\right) \hat{E}^{\left(+\right)}_p
    \hat{E}^{\left(-\right)}_s  \hat{E}^{\left(-\right)}_i  \left|0,0 \right>.
    \label{pdc-state}
\end{equation}
The susceptibility \(\chi^{\left(2\right)}\left(z\right)\) indicates the second-order nonlinearity, $V$ is the volume of the waveguide, and $t'$ denotes the interaction time. The three interacting fields \(\hat{E}_{\mu}(\mu = p, s, i)\) identify the pump field and the down-converted signal and idler waves, respectively. Note that we neglected the vacuum state which is of no interest in the scope of this paper.

The boundary conditions imposed on the fields propagating inside a waveguide define a discrete and finite set of allowed transverse field modes. This is in contrast to the infinite, continuous set of transverse modes possible in bulk crystal PDC. We derive the joint spatio-spectral wave function of the photon-pair by quantizing the fields of the signal and idler beams in the waveguide basis \cite{walls_quantum_1995}. 
For a fixed frequency we denote the spatial waveguide modes in terms of a discrete set of orthonormal mode functions  \(u_n(\bm{r}) \, (n = 1,\dots  ,N)\), where \(\bm{r}=(x,y)\) is the position in the transverse direction and $z$ is the
propagation direction of the pump beam. The vector \(\bm{q}_n=(k_x, k_y)\) indicates the transverse
momentum associated with the $n$th spatial mode of the field, and the reduced longitudinal wave numbers of pump, signal, and idler beams are given by 
\(\beta_n^{\left(p,s,i\right)}(\omega) = \sqrt{\left[\omega\, n(\omega)/c\right]^2 -
\left|\bm{q}_n\right|^2}\).
Their positive and negative frequency parts can now be written as:
\begin{equation}
\begin{split}
    \hat{E}_{\mu}^{(-)}(\bm{r},z,t) &= \hat{E}_{\mu}^{(+)\dagger}(\bm{r},z,t)\\&= 
    B' \sum_n \int d\omega_{\mu} u_n^{\left(\mu\right)}\left(\bm{r}\right)\\ &\times\exp \left\{\imath \left[\beta_n^{\left(\mu\right)}(\omega_{\mu})z - \omega_{\mu}  t\right] \right\}  \hat{a}^{\left(\mu\right) \dagger}_n(\omega_{\mu})
    \label{signal/idler field}
\end{split}
\end{equation}
where \(\mu = (s,i) \) labels the signal and idler fields and the sum embraces all spatial modes propagating inside the waveguide. In Eq. (\ref{signal/idler field}) all constants and slowly varying field amplitudes have been merged into the
overall parameter \(B'\), and the operator 
\(\hat{a}^{\left(\mu\right)\, \dagger}_n(\omega_{\mu})\) corresponds to the creation of one photon with a
given frequency in one discrete spatial mode.

If we treat the electric field of the strong pump beam classically with the approximation of an undepleted pump it can be written as:
\begin{equation}
\begin{split}
    E_p(\bm{r},z,t) = &\sum_{l} \int_0^{\infty} \mathrm d\omega \,
    \alpha(\omega)  A_l^{\left(p\right)}u_l^{\left(p\right)}\left(\bm{r}\right)\\
    &\times\exp\left\{\imath[ \beta_l^{\left(p\right)}(\omega) z - \omega t]\right\} + c.c. .
    \label{pump field}
\end{split}
\end{equation}
In the general case of ultrashort pump pulses the laser source can exhibit a broad-band frequency structure
which is modeled by the pump function \(\alpha(\omega)\). In the following the field amplitude of each pump 
spatial mode \(u_l^{\left(p\right)}\left(\bm{r}\right)\) is denoted by \(A_n^{\left(p\right)}\) (in the experiment this may be tuned by changing the coupling into the waveguide).

Using Eqs.(\ref{pdc-state})-(\ref{pump field}) and following Grice and Walmsley \cite{grice_spectral_1997} we find for the two-photon state:
\begin{equation}
\begin{split}
    \left|\psi\right> =  &B \sum_{lmn} A^{\left(p\right)}_l\iint_0^{\infty} \mathrm  d\omega_s\, \mathrm d\omega_i\, \\
    &\times \underbrace{\int_A \mathrm d\bm{r}\,    u_l^{\left(p\right)}\left(\bm{r}\right)  u_m^{\left(s\right)}\left(\bm{r}\right) u_n^{\left(i\right)}\left(\bm{r}\right)  }_{A_{lmn}}
   \alpha(\omega_s + \omega_i) \\
    &\times \underbrace{\mathrm{sinc}\left[\Delta \beta_{lmn}(\omega_s, \omega_i) L
    / 2\right] \exp\left[\imath \Delta \beta_{lmn}(\omega_s, \omega_i) L /
    2\right]}_{\phi_{lmn}\left(\omega_s,\omega_i\right)}\\&\times \hat{a}^{\left(s\right)\dagger}_m(\omega_{s})\hat{a}^{\left(i\right)\dagger}_n(\omega_{i})
    \left|0,0 \right>
    \label{two-photon state evaluated}
\end{split}
\end{equation}
with $B$ defining an overall constant.  The indices $lmn$ label the triplet of the pump mode $l$, signal mode $m$, and idler mode $n$.

In Eq. (\ref{two-photon state evaluated}) we introduced the value \(\Delta \beta_{lmn}(\omega_s,
\omega_i) = \beta_l^{\left(p\right)}(\omega_s + \omega_i) - 
    \beta_m^{\left(s\right)}(\omega_s) -
    \beta_n^{\left(i\right)}(\omega_i)-\beta_{QPM} \), which corresponds to a momentum mismatch between the different propagation constants. This is corrected by the quasi-phase-matching  vector \(\beta_{QPM} \) arising from a periodic variation in the \(\chi^{(2)}\) nonlinearity in the $z$ direction fabricated in the waveguide to achieve perfect phase-matching \cite{di_giuseppe_entangled-photon_2002}. The spectrum of the down-converted photon pairs is given by the joint spectral amplitude (JSA) \(f_{lmn}(\omega_s, \omega_i) = \alpha(\omega_s + \omega_i)\times \phi_{lmn}(\omega_s,\omega_i)\), where \(\alpha(\omega_s + \omega_i)\) is the pump distribution and \(\phi_{lmn}(\omega_s, \omega_i)\) is the phase-matching function.

\begin{figure}
\includegraphics[width=\linewidth]{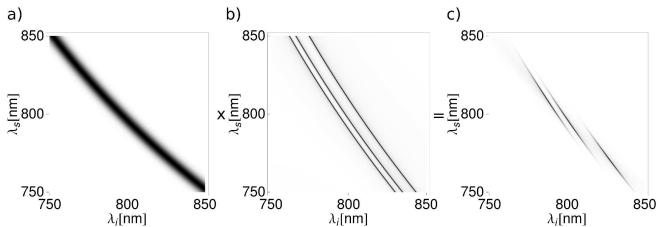}
\caption{\label{fig:multi-modeJsa} Different spatial modes excited in the
down-conversion process lead to a superposition of Gaussian shaped frequency
distributions in the generated two-photon state. (a) pump function, (b) phase-matching function, and (c) JSA.}
\end{figure}

    Note that Eq. (\ref{two-photon state evaluated}) is similar to the two-photon state of collinear, plane-wave PDC, neglecting the spatial structure of the propagating waves \cite{grice_spectral_1997}. However, by explicitly considering the spatial modes propagating inside the waveguiding material, we can see that the generated biphotonic state is emitted into a superposition of interacting mode triplets ($lmn$). Each triplet typically exhibits a different overall down-conversion generation efficiency due to the overlap between the three interacting fields \(A_{lmn}\). Moreover the triplet possesses a unique spectrum \(f_{lmn}(\omega_s, \omega_i)\) because of the different longitudinal wave vectors satisfying the phase-matching condition  \(\Delta \beta_{lmn}(\omega_s, \omega_i) = 0\). In the experiment the spread of the photons into different spatial modes can be tuned by controlling the preparation of the incoming pump wave into different waveguide modes \(A_l^{\left(p\right)}\). In conclusion the multi-mode PDC state is represented as:

\begin{equation}
\begin{split}
    \left|\psi\right> =  &B \sum_{lmn} A_l^{\left(p\right)} A_{lmn} \iint_0^{\infty} 
 \mathrm  d\omega_s\, \mathrm d\omega_i\,   f_{lmn}(\omega_s,\omega_i) \\&
 \times \hat{a}^{\left(s\right)\dagger}_n(\omega_{s})\hat{a}^{\left(i\right)\dagger}_m(\omega_{i})
    \left|0,0 \right>
    \label{two-photon state reduced}
\end{split}
\end{equation}
The JSA of a typical two-photon state with only three different mode triplets is plotted in Fig. \ref{fig:multi-modeJsa}. As can be seen the differences in the momentum mismatch lead to slightly shifted phase-matching functions and the photons are generated in a superposition of these three frequency amplitudes. Hence the down-converted photon pairs are emitted into a composite state entangled in frequency and spatial mode (or transverse momentum).

\section{\label{sec:model} Waveguide model}
To model numerically the properties of the generated two-photon state for a given waveguide architecture and down-conversion process, we assume a rectangular waveguide with width $W$, height $H$, a difference in the refractive index  between the  waveguide and substrate materials of $\Delta n$ and a poling period $\Lambda$. The waveguide dimensions and the difference in the index of refraction are deduced from a measurement of waveguide's numerical aperture. The extracted parameters suggest a
\begin{figure}
    \includegraphics[width=0.8\linewidth]{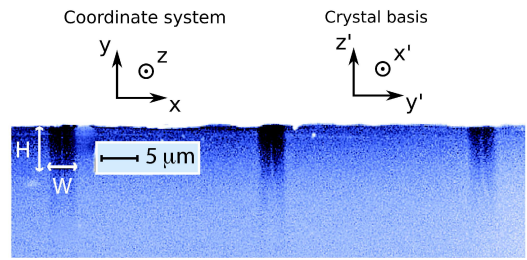}
    \caption{\label{fig:waveguidePic} (Color online) Microscope image of the waveguide cross-section. The height and the width ($H \times W= 6 \times4$ $\mu$m$^2$) given by the numerical aperture measurement are in good agreement with the image. Note the different nomenclature for the biaxial crystal basis and the coordinate system used for the calculations.} 
\end{figure}
rectangular waveguide with parameters $H = $ 6 \(\mu\)m, $W$ = 4 \(\mu\)m,  and $\Delta n $ = 0.01. In addition to the refractive index difference $\Delta n$ between the waveguide and the substrate, our model also takes into account the air boundary at the upper edge of the waveguide. The waveguides have an effective poling period of $\Lambda = 7.59\, \mu$m. The estimates for the waveguide dimensions were verified with a microscope as shown in Fig.~\ref{fig:waveguidePic}. 

\begin{figure}
    \includegraphics[width=0.8\linewidth]{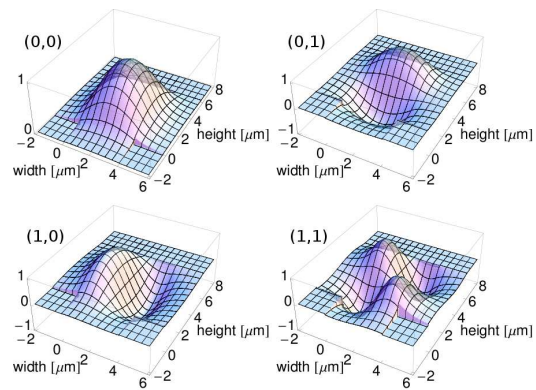}
\caption{\label{fig:fieldModesComp} (Color online) The first four $z'$ polarized spatial field modes at 800 nm propagating in our waveguide.}
\end{figure}

In our type-II PDC process $y'$ polarized pump photons near 400 nm are down converted into $y'$ and $z'$ polarized signal and idler photons around 800 nm.
The spatial modes propagating in this waveguide architecture can be calculated 
with a semi-analytical dielectric waveguide model \cite{marcuse_theory_1974}. The solutions of the transverse wave equations correspond to sine and cosine functions followed by an exponential decay into the surrounding material (see Fig. \ref{fig:fieldModesComp}).
The spatial modes and their respective transverse momenta can be conveniently labeled by the number of nodes 
in the transverse $x$ and $y$ directions: ($x$ nodes, $y$ nodes).  Our waveguide architecture shows multi-mode behavior at the spectral range of interest and supports spatial modes from (0,0) up to (3,5) and (1,2) at 400 and 800 nm, respectively.
For given field modes we can readily evaluate both  the propagation  constant  \(\beta_\nu^{\left(p,s,i\right)}\) ($\nu = l, m, n $) and the
coupling constant  \(A_{lmn}\)   
for each mode triplet in Eq.~(\ref{two-photon state evaluated}) (cf.~Table~\ref{tab:modeCoupling}).
\begin{table}
    \centering
   \begin{tabular*}{0.48\textwidth}{@{\extracolsep{\fill}} c c c c c r} \hline \hline
        \( E^{\left(p\right)}(l_x, l_y)\)  & \(\rightarrow\) & \(E^{\left(s\right)}(m_x, m_y)\)  & + & \(E^{\left(i\right)}(n_x, n_y)\) & \(A_{(lmn)} (10^{-4})\)\\ \hline
        (0,0)               & \(\rightarrow\) & (0,0)               & + & (0,0) & 53.96\\
        (0,0)               & \(\rightarrow\) & (0,1)               & + & (0,0) & 0.11\\
        (0,0)               & \(\rightarrow\) & (0,1)               & + & (0,1) & 35.66\\
        (1,0)               & \(\rightarrow\) & (1,0)               & + & (0,0) & 42.55\\
        (0,2)               & \(\rightarrow\) & (0,0)               & + & (0,0) & 4.72\\
        \hline \hline
    \end{tabular*}
    \caption{Five selected coupling constants for different down-conversion processes possible in our WG architecture.}
    \label{tab:modeCoupling}
\end{table}

In the special case of a rectangular waveguide the parity between the interacting modes is conserved  \( (l_{i} + n_{i} + m_{i} = 2N, N\in\mathds{N}  \)) similarly to the conversion of OAM in bulk crystal PDC. All down-conversion processes violating the parity conservation exhibit a vanishing coupling coefficient \(A_{lmn}\).
However perfect parity conservation would only be given if signal, idler, and pump waves propagate in the same mode basis, which in general is not the case due to their different polarizations and wavelengths. Hence weak coupling exists between mode triplets that do not conserve parity.

\begin{figure}
    \includegraphics[width=0.8\linewidth]{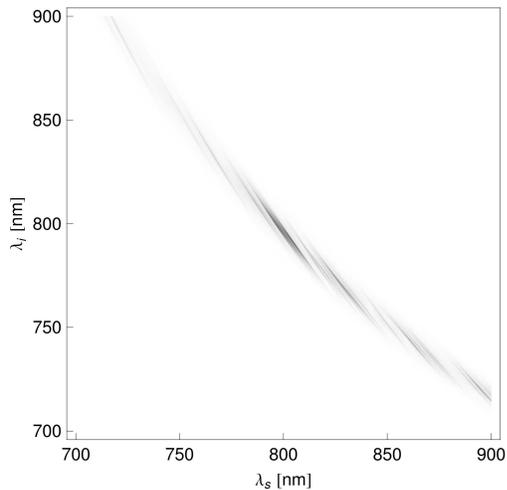}
    \caption{\label{fig:multi-modeJSA} Frequency distribution of the two-photon state for the 60 most important mode triplets. The PDC spectra are spread over a range of more than 200 nm.}
\end{figure}

Altogether we expect for this configuration that the down-converted fields will be emitted in a superposition of 720 different mode triplets each having a distinct JSA and coupling efficiency. With a 1.14 nm pump bandwidth the frequency spectra are spread over 200 nm in wavelength as can be seen in Fig.~\ref{fig:multi-modeJSA}.

\section{\label{sec:shg} Study of spatial structure by second harmonic measurements}
SHG is a useful tool to analyze the spectral effects of spatial-mode propagation in our waveguides because, unlike PDC, all three fields are intense. Hence it is easy to measure both their spectral and spatial characteristics. In this way we can anticipate the spatio-spectral structure of the PDC emission from the waveguide by studying its SH response. The rich modal structure of the potassium titanyl phosphate (KTP) waveguides has been studied in the past by Roelofs \textit{et al.}, who reported the existence of higher-order spatial modes by investigating frequency doubling \cite{roelofs_characterization_1994}.

\begin{figure}
\includegraphics[width=0.7\linewidth]{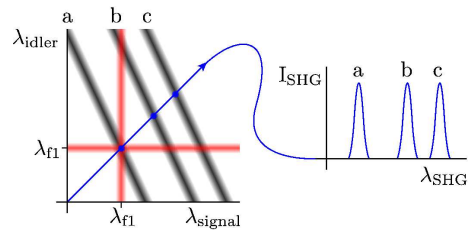}
\caption{\label{fig:shgSchematic} (Color online) Schematic of SHG. The two gray (red) lines represent the two pump fields and the three black curves are three different phase-matching functions for three mode triplets. The black (blue) line at 45$^{\circ}$ is the locus of points picked out by the pump function as the fundamental wavelength is scanned. SH signal can be generated at the intersections of this line and the three phase-matching functions.}
\end{figure}

In SHG, the signal and  idler modes at the fundamental frequency are combined to form a second-harmonic mode at double the frequency. Thus the wavelengths of pump and output are reversed and in analogy to PDC this process is phase matched at the same frequencies: an SHG signal will only be present at wavelengths where pump and phase-matching functions overlap. Tuning the pump (fundamental) frequency moves the pump function along a +45$^{\circ}$ line in the joint frequency space as shown in Fig.~\ref{fig:shgSchematic}. As this pump function crosses each individual phase-matching function a SH signal will be observed. If we use a pump with a sufficiently narrow bandwidth and align the pump beam such that we excite multiple waveguide modes, we can probe the position of the phase-matching functions for each spatial-mode triplet.
 
\begin{figure}
\includegraphics[width=0.7\linewidth]{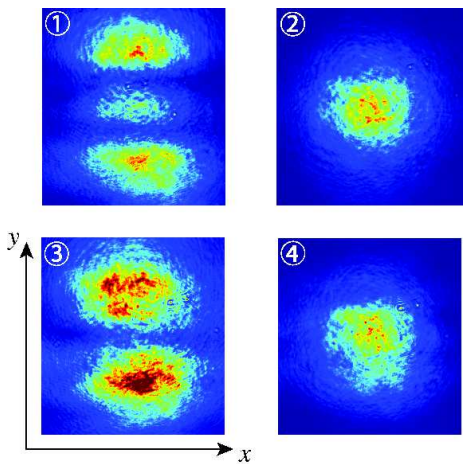}
\caption{\label{fig:shgModes} (Color online) Observed second-harmonic modes. The modes (1) (0,2), (2) (0,0), (3) (0,1) and, (4) (0,0) were recorded at wavelengths of 393.9 nm, 397.9 nm, 400.0 nm and 404.6 nm respectively. (Axis labels correspond to the coordinate system in Fig. \ref{fig:waveguidePic}.)}
\end{figure}

In our first experiment we pumped a 2.1-mm-long, type-II PP-KTP (periodically poled) waveguide with a narrow-band Ti:sapphire laser [0.6 nm full width at half maximum (FWHM), 2.6 ps autocorrelation length, and $\sim$1.5 MHz repetition rate] and measured the second-harmonic response with a spectrometer (Ocean Optics, Mikropack) while tuning the pump from 780 to 820 nm. Simultaneously, we recorded the spatial structure of the frequency doubled mode with a beam profiler charged-coupled device (CCD) camera (uEye, IDS Imaging).
We observed a SH response at  four different wavelengths (Table \ref{tab:shgPeaks}).
The corresponding spatial modes  can be identified as (0,2), (0,0), (0,1), and (0,0) as shown in Fig.~\ref{fig:shgModes} and the SH spectra of the different spatial modes lie 2--5 nm apart. These modes propagate in the waveguide and could therefore be exploited to pump down-conversion processes.
 
With the help of the SH signal we verified the results of our model (Sec.~\ref{sec:model}), specifically that it correctly predicts the separation between different SH peaks. In order to match exactly the absolute wavelength values from our model with the measured values, we introduced a small correction into the pump momentum. This was used as a fitting parameter to correct a discrepancy between the Sellmeier equations \cite{kato_sellmeier_2002} calculated for a rectangular waveguide and those for the real waveguide used in the experiment. It should be noted that the value of this correction was only around 0.4\% of the pump momentum and was the only free parameter in the model. The measured and predicted positions of the SH peaks were then in very good agreement as shown in Table \ref{tab:shgPeaks}. 

\begin{table}
    \centering
     \begin{tabular*}{0.48\textwidth}{@{\extracolsep{\fill}}c c c c c c c c} \hline \hline
      & &  &  & & & Experiment & Theory \\
   Mode &   \( E^{\left(f_2\right)}\)  &+ & \(E^{\left(f_1\right)}\)  &  \(\rightarrow\) & \(E^{\left({SHG}\right)}\)  & \(\lambda_{\textrm{SHG}}\)(nm) & \(\lambda_{\textrm{SHG}}\) (nm) \\
        \hline
 1      &  (0,0)       & + & (0,0)       & \(\rightarrow\) & (0,2)  & 393.9 & 395.1 \\
 2      & (0,0)       & + & (0,0)       & \(\rightarrow\) & (0,0)  & 397.9 & 398.0 \\
 3      & (0,0)       & + & (0,1)       & \(\rightarrow\) & (0,1) & 400.0 & 399.3 \\
4       & (0,1)       & + & (0,2)       & \(\rightarrow\) & (0,0)  & 404.6 & 404.1 \\ \hline \hline
    \end{tabular*}
    \caption{\label{tab:shgPeaks} Spectral correspondence of different spatial modes in SHG. The second column states the mode triplets that yield the recorded wavelengths. The third and fourth columns show the measured and predicted positions of the SH modes.}
\end{table}

\section{Measurement of spatial-to-spectral coupling in PDC}
In our next experiment we investigated the spatial structure of the signal and idler photons from PDC more directly by measuring the spectrum of the single counts in both signal and idler arms. 
Measuring the single counts of only one arm discards the events in the other. However the corresponding spectral peaks of the two measured down-conversion distributions can be connected by energy conservation. This type of spectral marginal measurement corresponds to an integration over one spectral dimension of the JSA.

\begin{figure}
\includegraphics[width=0.8\linewidth]{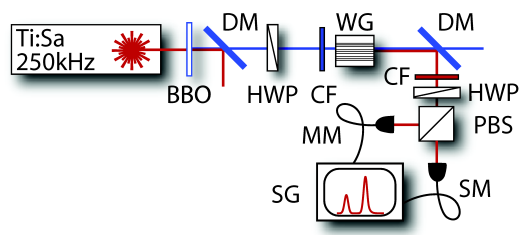}
\caption{\label{fig:pdcSetup} (Color online) Experimental setup for measuring the spectral marginals of the signal and idler. BBO, $\beta$-barium borate crystal for SHG; DM, dichroic mirror; CF, color glass filter; HWP, half-wave plate; WG, waveguide; PBS, polarizing beam splitter; MM, multi-mode fibre; SM, single-mode fibre; SG, spectrograph with CCD-Camera.}
\end{figure}

We pumped our down-converter with the SH of a mode-locked Ti:sapphire laser (798 nm, 10 nm FWHM, and 250 kHz repetition rate). 
A fairly narrow-band pump was required in order to resolve the spectral peaks arising in the marginal distributions from different spatial modes: hence the SH had a spectral width of 1.1 nm centered around 399 nm.  
Our nonlinear medium was a 3.5-mm-long type-II PP-KTP waveguide similar to the one studied in Sec.~\ref{sec:shg}. 
Signal and idler were separated at a polarizing beam splitter; one arm was coupled into a single-mode (SM) fiber, while the other was coupled into a multi-mode (MM) fiber. The fiber outputs were connected to a high-sensitivity spectrograph (Andor) as shown in Fig.~\ref{fig:pdcSetup}. By rotating a half-wave plate before the polarizing beam splitter we could direct either signal or idler beam to the MM or SM fiber.

\begin{figure}
\includegraphics[width=0.8\linewidth]{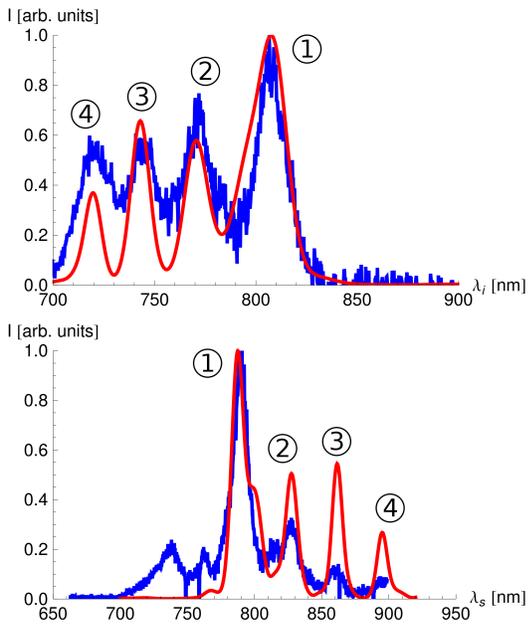}
\caption{\label{fig:marginalsTheorExp} (Color online) The observed noise-subtracted signal and idler marginals in black (blue) are in very good accordance with the 
predicted spectra in gray (red).}
\end{figure}

Coupling to MM fibers allowed the spectra of all of the different PDC spatial modes to be observed simultaneously. The large number of spectral peaks indicates that several spatial modes were excited inside the WG (Fig.~\ref{fig:marginalsTheorExp}). As expected, the peaks in the multi-mode signal and idler spectra could be accurately paired according to energy conservation. When coupling to SM fiber the main spectral peak corresponding to  (0,0) mode in signal (790 nm) and idler (806 nm) dominates indicating that only the fundamental spatial mode is efficiently coupled.

\begin{table}
    \centering
    \begin{tabular*}{0.48\textwidth}{@{\extracolsep{\fill}}c  c c c c c c c} \hline \hline
       Proc. & \( E^{\left(p\right)})\)  & \(\rightarrow\) & \(E^{\left(s\right)}\)  & + & \(E^{\left({i}\right)}\) & \(\lambda_{s}\) (nm) & \(\lambda_{i} \) (nm)\\
        \hline
        1 &   (0,0)       & \(\rightarrow\) & (0,0)       & + & (0,0)                & 788.1 & 806.2\\
        2 &   (0,0)       & \(\rightarrow\) & (0,1)       & + & (0,1)                & 827.0 & 700.3\\
        3 &   (0,0)       & \(\rightarrow\) & (1,0)       & + & (1,0)                & 816.5 & 742.9\\
        4 &   (0,0)       &  \(\rightarrow\)& (1,1)       & + & (1,1)                & 895.0 & 719.4\\ \hline \hline
    \end{tabular*}
    \caption{\label{tab:pdcPeaks} Spectral correspondence of different spatial modes observed in the marginal measurements of the PDC.}
\end{table}

A crucial factor in influencing the structure of the photon pairs is the coupling of the pump mode into the
waveguide. We examined the overlap between the incident free-space Gaussian pump mode
and the corresponding pump modes in the waveguide basis, \( A^{\left(p\right)}_l = \int_A \mathrm d\bm{r}\, u_l^{\left(p\right)}\left(\bm{r}\right) E^{\left(p\right)}_{ext.}\left(\bm{r}\right)\). For a Gaussian pump beam with the correct beam waist properly aligned with the center of the waveguide almost all of the energy is deposited into the (0,0) waveguide mode and very little is coupled into higher-order waveguide modes. Nevertheless given that small pump misalignments were present we estimated a relative coupling into each higher order pump mode of 10\% of the energy coupled into the (0,0) mode. Taking into account this coupling of the beam into the waveguide in conjunction with the conversion efficiency of all the interacting mode triplets and their respective phase-matching functions, we calculated the generated JSA. Hence we were also able to predict the measured marginal spectra with very high accuracy. We observed that the output is dominated by four main down-conversion processes (Table \ref{tab:pdcPeaks}); the coupling to other possible down-conversion processes is suppressed such that they cannot be seen above the noise level in the measurement. The remaining small deviations from the predicted marginal measurements can be explained by the different coupling efficiencies of the various modes into the MM fibers and the differences between the real waveguide investigated and our rectangular waveguide model. Note that the predicted spectra of the SHG and the PDC measurements are in very good agreement with each other. Our model requires only one fitting parameter which is the small constant offset in the effective Sellmeier equations for the waveguide.

\section{Discussion}
Our results from the previous sections demonstrate that higher-order spatial mode propagation has significant consequences for photon-pair generation in waveguide architectures. The majority of quantum optics experiments require single photons in a well-controlled single mode in any given degree of freedom \cite{eckstein_broadband_2008, martin_integrated_2009}. 
Consequently, in most single photon experiments, it is desirable to suppress all higher-order spatial-mode processes and to promote the emission of the photon pairs into the fundamental mode [\( (0,0) \rightarrow (0,0) + (0,0)\)].

In order to achieve this with waveguided PDC, we first turn our attention to the measures one can take to promote the coupling of the pump beam into the fundamental waveguide mode. The pump beam should be spatially filtered to have a Gaussian profile, either by the use of a pinhole and lens pair or a single-mode fiber. This mode should be carefully matched to the waveguide diameter either by choosing the appropriate lenses for efficient coupling into the waveguide or by directly butt-coupling a single-mode fiber tip to the waveguide. In general a spot size with a diameter larger than the waveguide profile provides pump propagation that is closer to single mode, whereas a smaller beam will tend to exhibit more multi-mode effects. Fine tuning the coupling of the pump beam into the waveguide has a large impact on the modal structure of the pump and the optimum position can be found by monitoring the PDC spectrum. In our experiments we have found that this careful pump coupling can very effectively suppress all down-conversion processes originating from higher-order pump modes.

\begin{figure}
\includegraphics[width=0.8\linewidth]{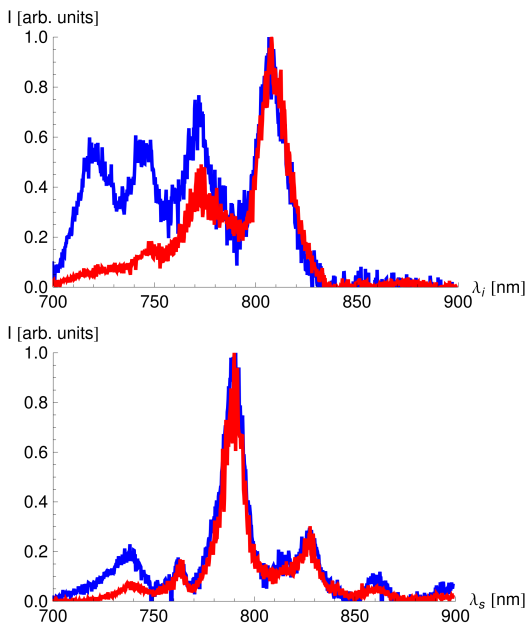}
\caption{\label{fig:marginalsSMMM} (Color online) Noise subtracted SM fiber measurement (gray, red) and MM fiber (black, blue) marginal measurements. In these graphs the impact of spatial filtering on the spectral measurements  can clearly be seen. The SM fibers select the signal and idler photons in the ground mode and suppress the side peaks from higher order spatial mode signal and idler photons.}
\end{figure}

Having implemented these changes, any spectral peaks observed in the output spectrum beyond those of the fundamental modes originate from the down-conversion of the pump photons from the fundamental mode into higher-order signal and idler modes, as already shown in Fig. \ref{fig:marginalsTheorExp}. Due to the different spectra connected to the different spatial modes, the output can be filtered in the spatial \emph{or} the spectral domain to ensure collection of emission from the fundamental signal and idler  mode only.
By coupling the output beams into single-mode fibers one can affect a spatial-mode filter with relatively high contrast. The impacts on the frequency spectrum of such a spatial mode filter can clearly  be seen in Fig. \ref{fig:marginalsSMMM} where the coupling of the down-converted photons was changed from MM to SM fiber. This demonstrates that the spectra of the beam can be manipulated by operations in the spatial domain. On the other hand, a frequency filter may be used to operate on the spatial structure of the photon pairs, and could be adjusted to only transmit photons with spectra originating from fundamental mode propagation.

Alternatively this problem can be addressed in the production process by reducing the size of waveguide cross section to only guide signal and idler beams in the fundamental mode. However since the wavelength of the down-converted pairs is approximately twice that of the pump, the light inevitably propagates not only in the fundamental waveguide mode but also in several higher-order spatial modes. Therefore multi-mode propagation effects still have to be taken into account, yet the number or the excited different down-conversion processes will be significantly reduced.

It is conceivable that the high down-conversion efficiency possible with a waveguide source in addition to these additional degrees of freedom may be used to create a multiplexed photon-pair source by simple spatial or spectral filter operations on the  generated biphotonic states.  On the other hand these modes could be utilized as an information-processing degree of freedom and hence achieving control over the spatial emission profile is of paramount importance. For example, the spatial correlations may be applied in the generation of hyperentangled states offering new possibilities in the production of Bell states and additional robustness in quantum error correction due to the enlarged Hilbert space.

\section{Conclusion}
In this paper we have derived the two-photon state generated in waveguided parametric down-conversion, explicitly taking into account the spatial mode propagating inside the waveguide material. We have shown how the propagation of a multitude of pump, signal, and idler modes affects the spatial \textit{and} spectral structure of photon pairs emitted from a nonlinear waveguide. We have observed these effects both through measurements of the spectra of second-harmonic light generated in a KTP waveguide as well as in marginal measurements of down-converted photon pairs. The spectra derived from the multi-mode treatment of the SHG and the down-conversion fit the measured data extremely well, yielding the possibility to precisely engineer further experiments. Furthermore we have shown that we can influence the spectra of the generated biphotonic states by operations in the spatial domain.

We suggested how these effects may be controlled to give a truly single-mode photon source in the spatial domain by either careful experimental design or waveguide engineering. These findings are of significance for quantum information experiments where nonlinear waveguides are used to generate photon pairs and may be harnessed to generate hyperentangled photon states or may be applied to design  multiplexed photon-pair sources.

\textit{Note added:}
Recently we became aware of an independent theoretical examination of spatial entanglement in waveguided PDC \cite{saleh_modal_2009}.
Furthermore a recent experimental investigation of higher order spatial modes in KTP waveguides with sum frequency generation was brought to our attention \cite{karpinski_experimental_2009}.

\section*{Acknowledgement}
This work was supported by the EC under the FET-Open grant agreement CORNER, (Grant No. FP7-ICT-213681).


\end{document}